\documentclass{aa}
\usepackage{graphicx, natbib,epsfig,amsmath,amssymb, mathrsfs, txfonts}
\DeclareGraphicsExtensions{.eps, .jpg, .ps, .png}
\bibpunct{(}{)}{;}{a}{}{,}
\begin{document}

\title{Design, analysis and test of a microdots apodizer for the \\ Apodized Pupil Lyot Coronagraph}
\author{P. Martinez\inst{1, 2, 5} \and C. Dorrer\inst{3}  \and E. Aller Carpentier\inst{1} \and M. Kasper\inst{1} \and A. Boccaletti\inst{2,5} \and K. Dohlen\inst{4} \and N. Yaitskova\inst{1}} 
\institute{European Southern Observatory, Karl-Schwarzschild-Strasse 2, D-85748, Garching, Germany 
\and LESIA, Observatoire de Paris Meudon, 5 pl. J. Janssen, 92195 Meudon, France 
\and Laboratory for Laser Energetics-University of Rochester, 250 East River Rd, Rochester, NY, 14623-USA 
\and LAM, Laboratoire d\'{}Astrophysique de Marseille, 38 rue Fr\'{e}d\'{e}ric Joliot Curie, 13388 Marseille cedex 13, France 
\and Groupement d\'{}int\'{e}r\^{e}t scientifique PHASE (Partenariat Haute r\'{e}solution Angulaire Sol Espace)} 
\offprints{P. Martinez, martinez@eso.org}

\abstract
{Coronagraphic techniques are required to detect exoplanets with future Extremely Large Telescopes. One concept, the Apodized Pupil Lyot Coronagraph (APLC), is combining an apodizer in the entrance aperture and a Lyot opaque mask in the focal plane. This paper presents the manufacturing and tests  of a microdots apodizer optimized for the near IR.}
{The intent of this work is to demonstrate the feasibility and performance of binary apodizers for the APLC. 
This study is also relevant for any coronagraph using amplitude pupil apodization.} 
{A binary apodizer has been designed using a halftone dot process, where the binary array of pixels with either 0\% or 100\% transmission is calculated to fit the required continuous transmission, i.e. local transmission control is obtained by varying the relative density of the opaque and transparent pixels. An error diffusion algorithm was used to optimize the distribution of pixels that best approximates the required field transmission. The prototype was tested with a coronagraphic setup in the near IR.}
{The transmission profile of the prototype agrees with the theoretical shape within 3\% and is achromatic.  
The observed apodized and coronagraphic images 
are consistent with theory. However, binary apodizers introduce high frequency
noise that is a function of the pixel size. Numerical simulations were used to specify pixel size in order to minimize this effect, and validated by experiment.}
{This paper demonstrates that binary apodizers are well suited for being used in high contrast imaging coronagraphs. The correct choice of pixel size is important and must be adressed considering the scientific field of view.}


\maketitle


\section{Introduction}
Direct detection and characterization of faint objects around bright astrophysical sources is challenging due to the large flux ratio and small angular separation. For instance, self-luminous giant planets are typically $10^{6}$ times fainter than the parent star in the near-infrared. Even higher contrasts of up to $10^{10}$ are needed to reach the realm of mature giant or telluric planets.
In order to achieve these contrast levels, dedicated instruments for large ground-based telescopes such as SPHERE or GPI  \citep{2006Msngr.125...29B,2006SPIE.6272E..18M}, or EPICS \citep{EPICS} for the future European-Extremely Large Telescope (E-ELT) will use powerful Adaptive Optics (extreme AO or XAO) systems coupled with coronagraphs. 

While the XAO system corrects for atmospheric turbulence and instrument aberrations, the coronagraph attenuates the starlight diffracted by the telescope in the image plane.
Since the invention of the stellar Lyot coronagraph \citep{1939MNRAS..99..580L},  
there has recently been impressive progress in the field leading to a wealth of different coronagraphs that can be divided into different families. In particular, the Apodized Pupil Lyot Coronagraph (APLC) \citep{2002A&A...389..334A, 2003A&A...397.1161S} appears to be well suited for ELTs and has been studied theoretically \citep{2005ApJ...618L.161S, 2007A&A...474..671M}. 
The APLC features amplitude apodization in the entrance aperture to reduce diffraction and a small Lyot mask in the focal plane.
It is the baseline coronagraph for e.g. SPHERE, GPI, and the Lyot Project \citep{2004SPIE.5490..433O}. \citet{Corono} further show that the APLC is also well suited to be used with ELTs considering their particular pupil shapes and segmented mirrors. 

A major issue with the APLC (and other coronagraphs using apodization such as is the dual zone coronagraph, \citet{2003A&A...403..369S}) is the manufacturing of the apodizer itself. So far, 
three concepts have been explored to manufacture apodizers: 1/ a metal layer of spatially variable thickness, 2/ electron-sensitized HEBS glass (high-electron beam sensitive glass), and 3/ an array of opaque pixels with spatially variable density.
The third concept has several advantages over the first and second ones. It is intrinsically achromatic and avoids wavefront phase errors introduced by a metal layer of variable thickness or the process of writing a HEBS pattern. 
Simplicity and reproducibility of the technique is also a major advantage.

A binary design using halftone dot process can be in principle generalized to any apodizer masks (APLC, Dual zone) and even conventional pupil apodization coronagraph \citep{1964J, 2001ApJ...548L.201N, 2005A&A...434..785A} or as an alternative manufacturing solution for binary shaped pupil coronagraph masks 
\citep{2003ApJ...590..593V, 2003ApJ...582.1147K, 2004ApJ...615..555V, 2007A&A...461..783E, 2008A&A...480..899E}.

In this paper, we report on the development (design and laboratory tests) of a binary apodizer for the APLC using a halftone dot process.
First we describe the binary mask principle and the algorithm used to distribute pixels across the pupil to best fit the required field transmission (Sect.\ref{principle}). Optimization of the design through pixels size is discussed in Sect. \ref{design} while in Sect. \ref{labo} we report on laboratory results obtained with a prototype using a near-IR bench which reproduces the Very Large Telescope (VLT) pupil. Finally, we conclude on the suitability of this technique for planet finder instruments in Sect. \ref{conclu}.

\section{Principle of microdots apodizer}
\label{principle}
A binary apodizer is made of an array of opaque pixels (i.e dots) on a transparent substrate. It is fabricated by lithography of a light-blocking metal layer deposited on a transparent glass substrate. Spatially variable transmission is obtained by varying pixel density.
An error diffusion algorithm was used to calculate the density distribution that best fits the required field transmission \citep{floyd, Ulichney, 2007JOSAB..24.1268D}. This algorithm chooses the transmission of a given pixel of the apodizer (either 0$\%$ or 100$\%$) by comparing the transmission required at this location to a 50$\%$ threshold, i.e. the transmission is set to zero if the required transmission is smaller than 50 $\%$, and to one otherwise (see Fig. \ref{apodT}). The induced transmission error is diffused to adjacent pixels that have not been processed yet by biasing the transmission required at the corresponding locations. This locally cancels the error of the binary optics relative to the required transmission. Such procedure is used for gray-level reproduction with black-and-white printing techniques \citep{Ulichney}. We inform the reader, that further details on the algorithm principle are presented in \citep{2007JOSAB..24.1268D}.

Shaping of coherent laser beams has also been demonstrated \citep{2007JOSAB..24.1268D} using this technique. The error diffusion algorithm has the advantage that the introduced noise is blue, i.e., the noise spectral density is only significant at high spatial frequencies. This allows the accurate generation of gray levels and rapidly varying shaping functions. In the specific case of the design of a coronagraph, the algorithm allows us to well match the PSF of the binary apodizer to the required apodized PSF up to a certain radial distance which could be chosen as the control radius of the AO system. In theory, better shaping results are obtained with smaller pixels \citep[i.e sampling problem,][]{2007JOSAB..24.1268D}, since this allows finer control of the local transmission and pushes the binarization noise to higher frequency. This will be further discussed in Sect. \ref{design}.
%

\section{Design optimization}
\label{design}

\begin{figure}[!ht]
\begin{center}
\includegraphics[width=4cm]{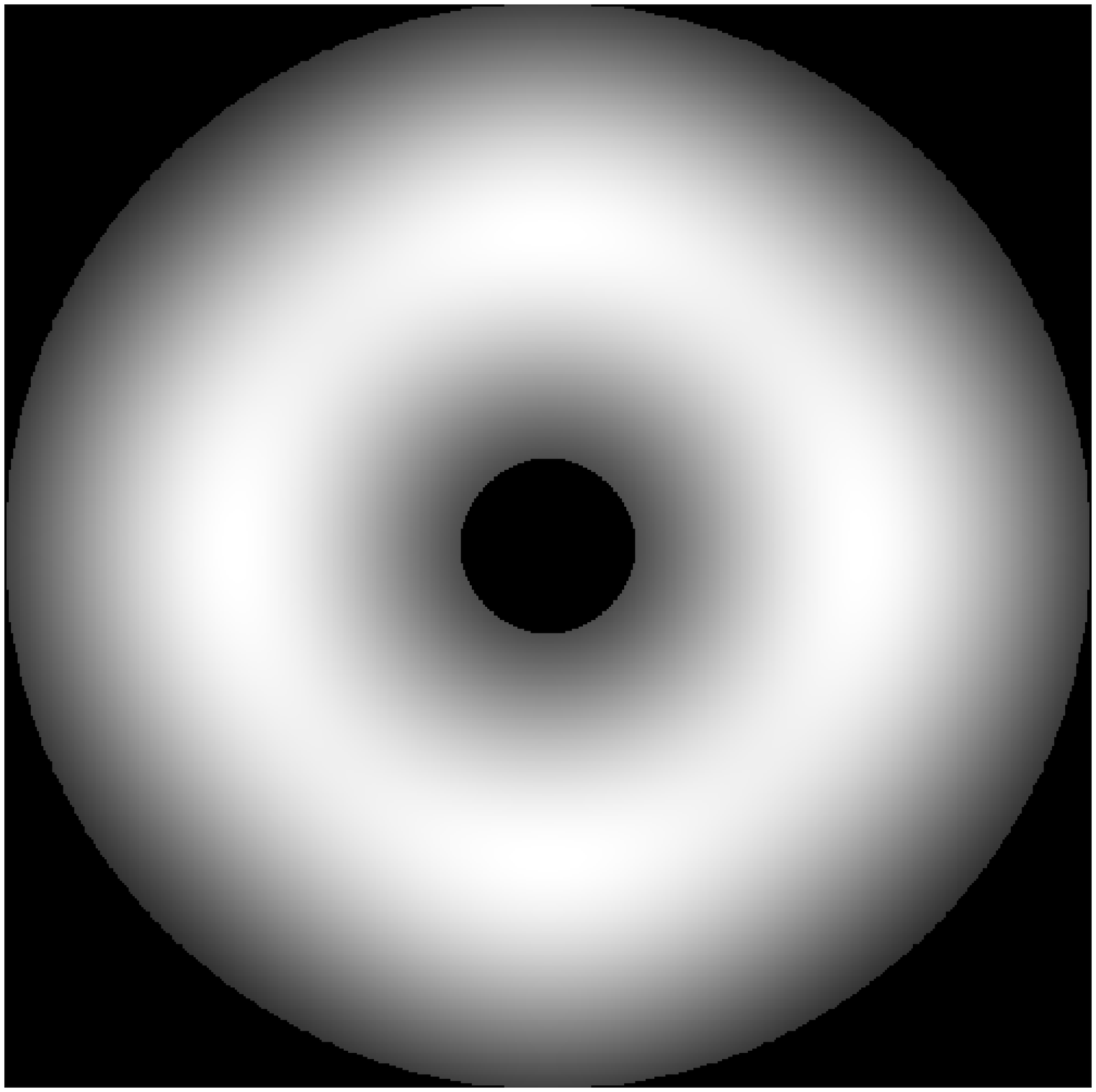}
\includegraphics[width=4cm]{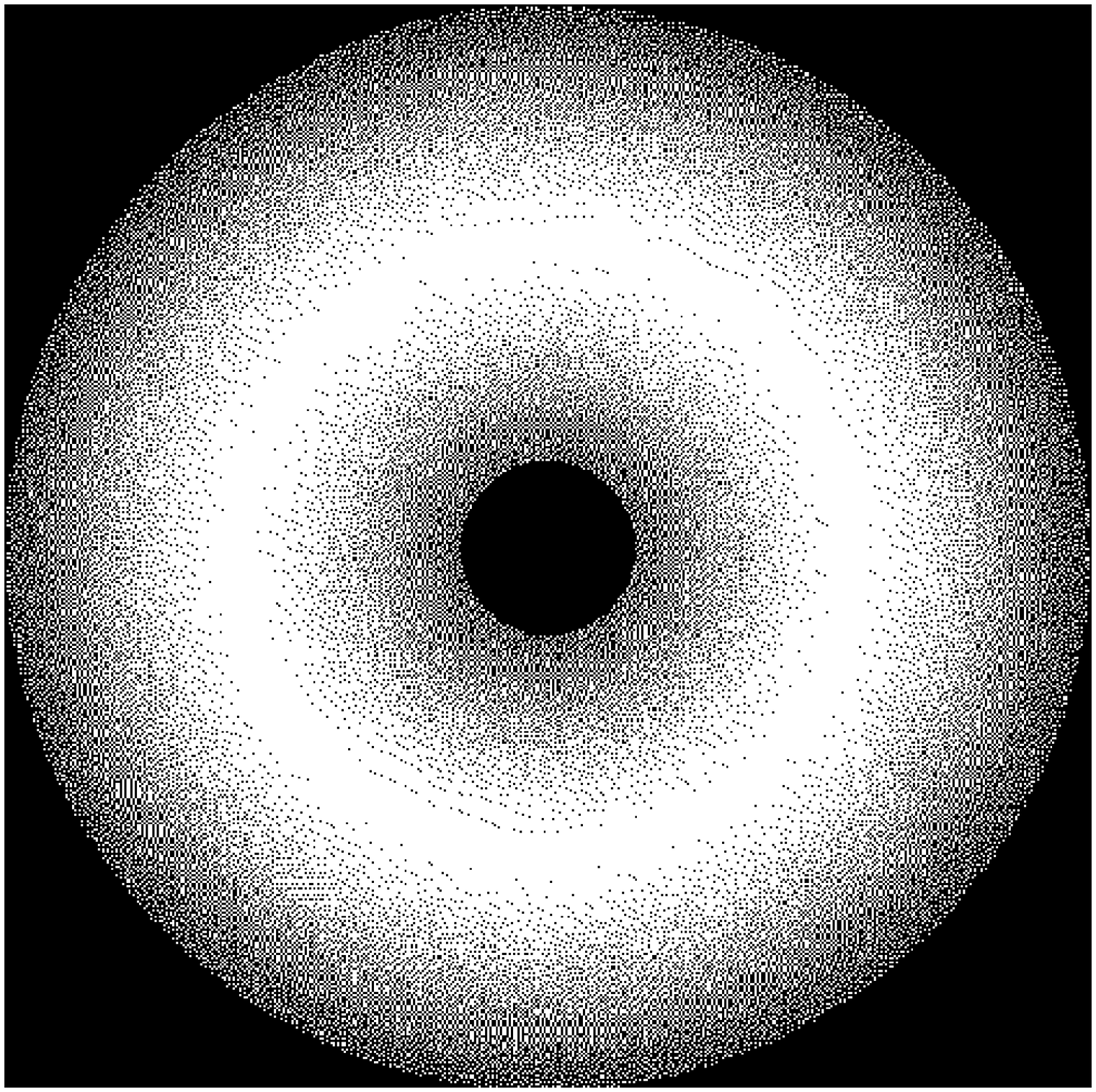}
\includegraphics[width=8.15cm]{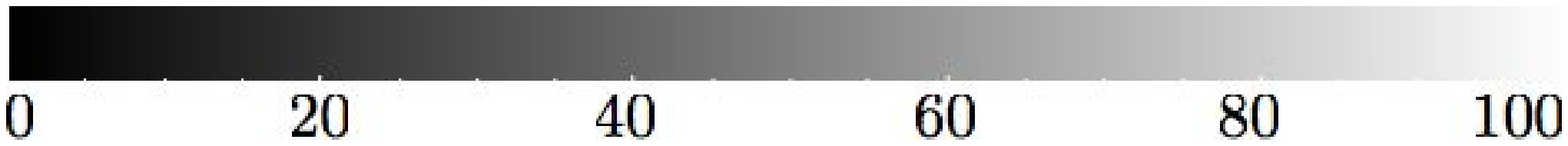}
\end{center}
\caption{Left: Shaper target (continuous apodizer). Right: Resulting microdots pattern using algorithm discussed in Sec. \ref{principle}. The spatial scale of these maps is 600$\times$600 pixels. The scale of transmission is given in $\%$.} 
\label{apodT}
\end{figure}

Assuming a VLT-like pupil, the apodizer is defined for a 15$\%$ central obscuration pupil  \citep[bagel regime,][]{2005ApJ...618L.161S}.
We consider a 4.5$\lambda/D$ APLC \citep{2007A&A...474..671M}. The apodizer shape is illustrated in Fig. \ref{apodT} (left image).
The inner-working angle of such configuration is $\sim 2.3 \lambda/D$.

\begin{figure}[!ht]
\begin{center}
\includegraphics[width=9cm]{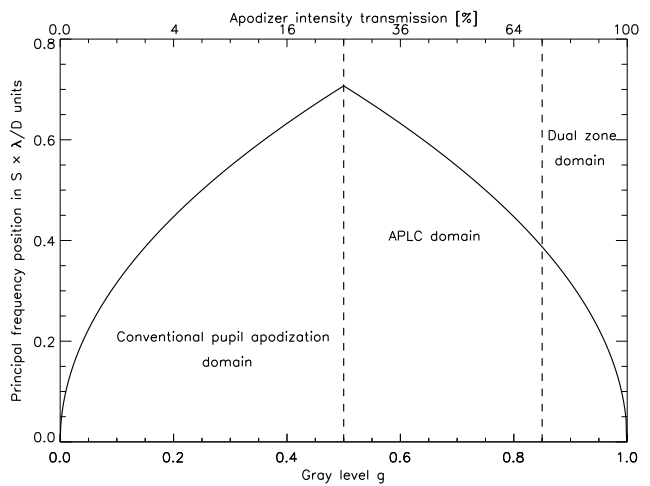}
\end{center}
\caption{First order peak diffraction $f_g$ position in $S \times \lambda/D$ units as a function of gray level $g$. Typical domain of application of apodizer masks are reported on the plot.} 
\label{summary}
\end{figure}  
The manufactured apodizer has a diameter of 3mm due to constraints on our optical bench (Sect. \ref{manufacturing} and \ref{labo}). 
For microdots, the performance is related to the ratio of the smallest feature 
to the pixel size. Hence, for the sake of clarity, we denote by $S$ the scaling factor, the ratio between the apodizer useful diameter (i.e pupil diameter, denoted $\Phi$ hereafter) and the pixel spacing, i.e pixel size (dot size), denoted $p$ hereafter:
\begin{equation}
S = \frac{\Phi}{p}
\label{scaling}
\end{equation} 

The individual pixels of a binary apodizer scatter light towards spatial frequencies depending on the pixel size. The smaller the pixels are, the higher are the spatial frequencies at which the light is scattered, and the better the achieved transmission profile matches the desired one. 

We also note that the high-frequency noise might have different distributions at different wavelengths. This would be a situation similar to diffraction gratings, where only diffracted orders (i.e corresponding to large values of the transverse wavevector $k$) are frequency-dependent. 
For such finer analysis, Fresnel propagators and a thorough modeling of the binary shaper (including process errors on the shape and size of each dot such as edge effects resulting from the isotropic wet etching process, Sect. \ref{manufacturing})   would be mandatory. 
\begin{figure*}[!ht]
\begin{center}
\includegraphics[width=9cm]{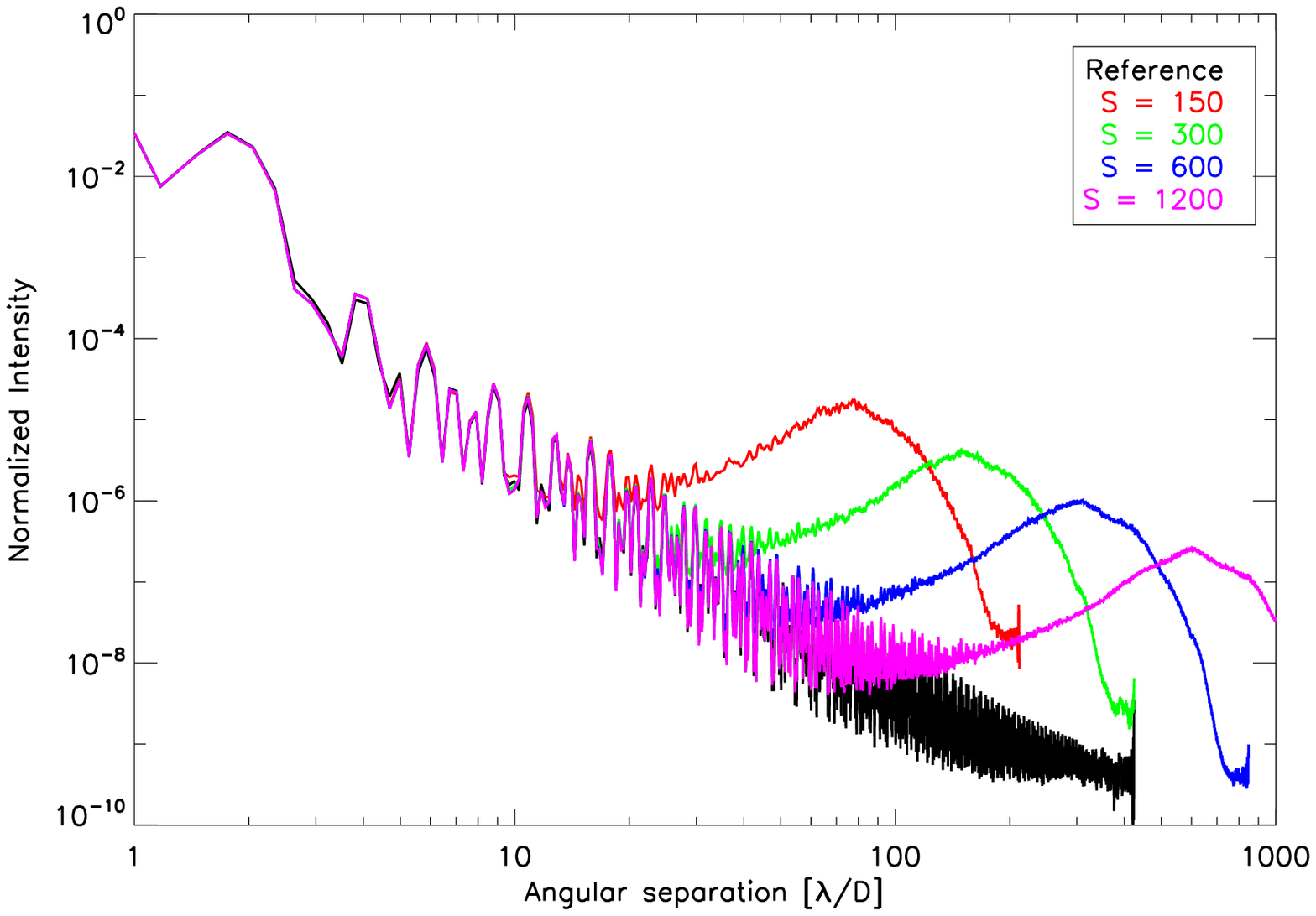}
\includegraphics[width=9cm]{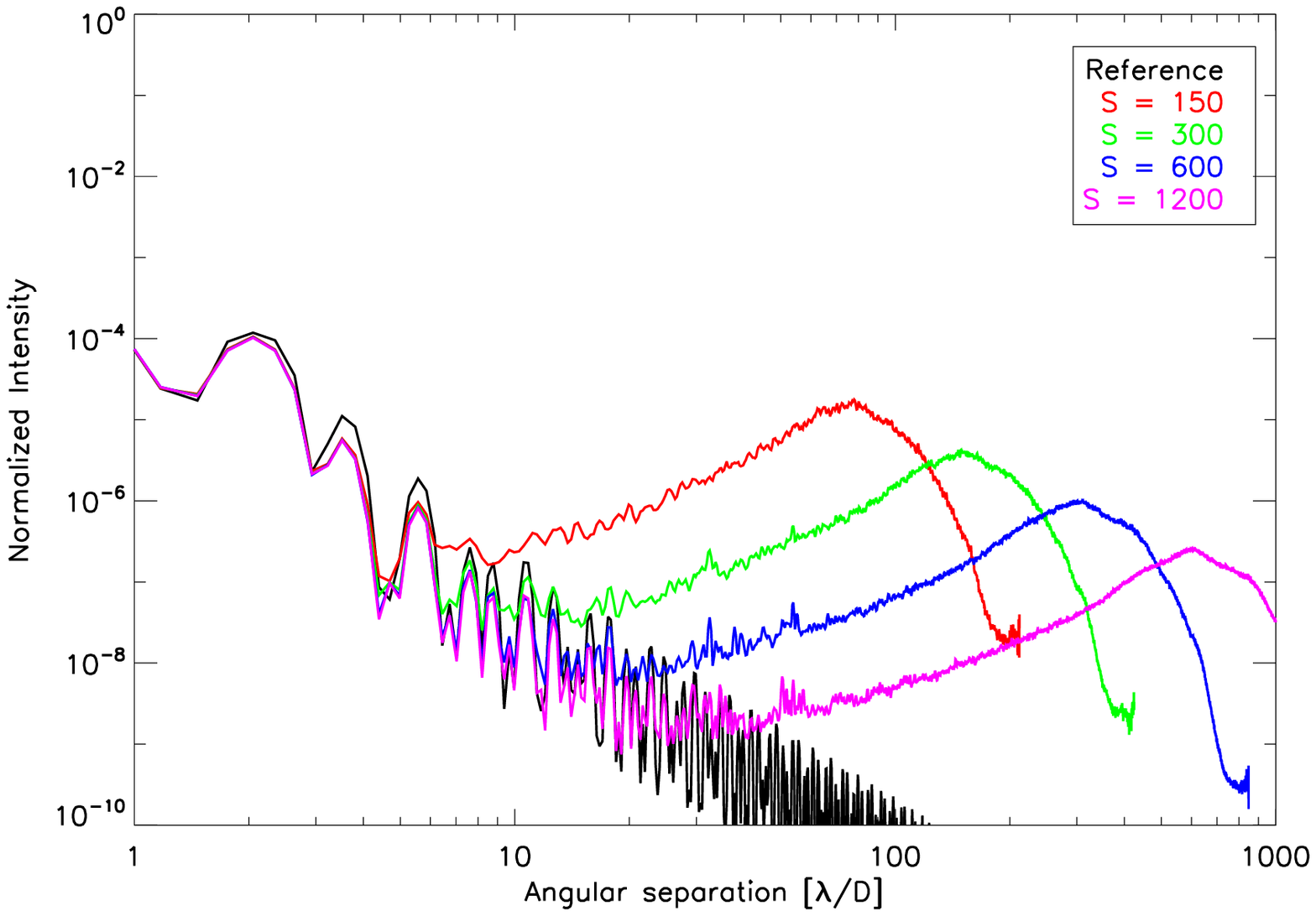}
\end{center}
\caption{Apodized PSFs (left) and APLC coronagraphic PSFs (right) using several dots size for the binary apodizer compared to that with continuous apodizer (i.e theory, in black). It assumes a pupil with 15$\%$ central obscuration. Profiles presented are azimuthal averages.}
\label{pixelsize}
\end{figure*} 
\subsection{Microdots diffraction stray light}

\noindent The microdots apodizer is modeled as an aperiodic under-filled two-dimensional grating which exhibits blue noise properties because of the error diffusion algorithm used \citep{Ulichney, 2007JOSAB..24.1268D}. 
The binary pattern produces an averaged gray level ($g = \sqrt{T}$, i.e averaged amplitude transmission) from an apodizer profile with intensity transmission $T$.
The resulting pattern spectral energy is set by $g$ (i.e by the minority pixels present on the device: non-metal pixels when $g < 0.5$ and by metal pixels conversely).
The spectral energy will increase as the number of minority pixels increases, peaking at $g=0.5$ \citep{Ulichney, Ulichney2}.
Most of the energy in the power spectrum of the pattern will be concentrated around the first order diffraction which would appear in the field of view at the spatial frequency $f_g$ (in  $\lambda/D$ units):
\begin{equation}
f_g =
\begin{cases}
\sqrt{g}\times S  & g \leq 1/2 \\
\sqrt{(1-g)}\times S  & g > 1/2 \\
\end{cases}
\label{principalF}
\end{equation}
\noindent Therefore, for a given $g$, the pattern power spectrum has a peak diffraction at $f_g$  \citep{Ulichney, Ulichney2}. As the gray level, $g$, increases from 0 to 0.5, the peak diffraction moves to further angular distance (Fig. \ref{summary}) with an increase of energy.
Above $g = 0.5$, situation is similar to (1-g), minority pixels has only changed from non-metal dots to metal dots. 
The PSF of a microdots device can be therefore expressed as function of a deterministic effect (the first order diffraction peak) bordered by speckles by stochastic effect (i.e dots distribution is not regular).
Higher order diffraction peaks are not relevant since out of the sciences field of view.
The intensity of the first order diffraction peak in the final coronagraphic image is function of $g$ as well.
The model presented hereafter is based on a study performed by \citet{Dust}, where effects of micro-obscurations such as dusts or cosmetic errors are analytically described for the SPHERE instrument image quality. 
The coronagraphic halo intensity ($I$) of the first order peak diffraction for N dots normalized to the stellar peak intensity is $ N_{dots} \times \left(\frac{p}{\Phi}\right)^{4}$ \citep[][assuming halos from all the dots add incoherently]{Dust}.
$N_{dots}$ is the total number of the minority dots present in the pattern and can be easily calculated through the surface ratio of the pupil by a square dot, times the density of minority dots ($\eta$, hereafter). 
The minority dots density can be expressed as:
\begin{equation}
\eta = 
\begin{cases}
g & g \leq 1/2 \\
1-g & g > 1/2 \\
\end{cases}
\label{density}
\end{equation}

\noindent then $N_{dots}$ is:
\begin{equation}
N_{dots} = \eta \times \frac{\pi}{4} \times \left(\frac{\Phi}{p}\right)^{2}
\end{equation}
The resulting relative halo intensity is then:
\begin{equation}
I = \eta \times \frac{\pi}{4} \times \left(\frac{1}{S}\right)^{2}
\label{I2}
\end{equation}
therefore, using Eq. \ref{density} one finally obtains:
\begin{equation}
I = 
\begin{cases}
g \times \frac{\pi}{4} \times \left(\frac{1}{S}\right)^{2} & g \leq 1/2 \\
(1-g) \times \frac{\pi}{4} \times \left(\frac{1}{S}\right)^{2} & g > 1/2 \\
\end{cases}
\label{I3}
\end{equation}

Considering our APLC apodizer ($T = 51 \%$, $g = 0.71$), the first order diffraction peak would be therefore localized at $f_g \sim S/2$ in $\lambda/D$ units with an intensity of $I \sim 1/(4\times S^{2})$.
For the Dual Zone coronagraph ($T \sim 80\%$, therefore $g \sim 0.9$), the first order diffraction peak moves closer to the central core of the PSF while its intensity decreases with respect to the APLC case:
$f_g \sim S/3$ in $\lambda/D$ with an intensity of $I \sim 1/(13\times S^{2})$.
For Conventional pupil apodization ($T \sim 25\%$, hence $g \sim 0.5$), the first order diffraction peak moves further away from the central core of the PSF while its intensity increases:
$f_g \sim 1/\sqrt{2} \times S$ in $\lambda/D$ with an intensity of $I \sim 2/(5\times S^{2})$.

\begin{center}
\begin{table*}
\centering
\begin{tabular}{c|c|c|c|c|c}
\hline \hline 
$S$ & $p$ [$\mu$m] & \multicolumn{2}{c|}{High frequency noise angular position [$\lambda/D$]} & \multicolumn{2}{c}{Microdots halo intensity} \\
\cline{3-6} 
& & Apodized PSF & Coronagraphic PSF & Theory ($I$) & simulation \\
\hline
150 & 20 & 20 & 5 & $1.0\times 10^{-5}$ &$1.7\times 10^{-5}$ \\
300 & 10 & 30 & 10 & $2.6\times 10^{-6}$ & $4.2\times 10^{-6}$\\
600 & 5  & 40 & 20 & $6.5\times 10^{-7}$ &$1.0\times 10^{-6}$ \\
1200 & 2.5 & 50 & 40 & $1.6\times 10^{-7}$ & $2.6\times 10^{-7}$\\
\hline
\end{tabular}
\caption{Angular position where the high frequencies noise appears on the apodized PSF and coronagraphic image as function of the pixel size (column 3 and 4). Microdots halo intensity as function of the pixel size: comparison between simulation (measured on the halo peak) and analytical expression $I$ (column 5 and 6). Results presented refers to Fig. \ref{pixelsize}.}
\label{table1}
\end{table*}
\end{center}

\subsection{Numerical simulations}
Our simulations make use of Fraunhofer propagators between pupil and image planes, which is implemented as fast Fourier transforms (FFTs) generated with an IDL code.
In the following, pixels describe the resolution element of the simulation, while dots describe the physical units forming the apodizer.
We use 0.3 $\lambda/D$ per pixel, while dots are sampled by 4 pixels to allow enough field of view to image the $1^{st}$ order diffraction peak for each $S$. 
Validity of the numerical dot sampling has been verified by comparing simulations with different dot sampling (1, 4 and 16 pixels per dot for $S = 150$).

We first analyze how the dot size affects the apodized PSF (Fig. \ref{pixelsize}, left) and the coronagraphic PSF (Fig. \ref{pixelsize}, right) with respect to an ideal continuous apodizer. From the results summarized in Table \ref{table1} we derive the following conclusions: 
\begin{itemize}
\item Reduction of the dot size by a factor of 2 increases the radial distance corresponding to an adequate agreement with the specification by a factor of 2 for the coronagraphic image. Eq. \ref{principalF} is confirmed by simulation.
\item Analytical model (Eq. \ref{I3}) is consistent with simulation predictions. This model is representative for the APLC situation. 
\item At a given frequency, in the coronagraphic images, the level of the noise decreases proportional to $S^4$ (for instance, at 80$\lambda/D$ noise increases from $3.2\times10^{-9}$ to $3.5\times10^{-5}$ when increasing $S$ by a factor 8) 
\end{itemize}

In practice, for the selection of $S$ (dots size), we modeled in simulation our specific pupil (VLT-like including the secondary support, i.e contrast accessibility issue) and taking into account the field of view of interest (sets by the AO correction domain: 20$\lambda/D$ like in SPHERE).
We found that as expected, the radial distance corresponding to an adequate agreement with the specification (ideal model) moved to 
larger angular separations while the intensity level where the noise appears remains in the order of the previous case.
In our context, $S=600$ ($5\mu m$ dots) meets our specifications.
The $S = 1200$ configuration leads to really small dot size (2.5$\mu$m). In such a case, when the dot size is of the order or lower than the operating wavelength (1.65$\mu$m for our application) a Rigorous Coupled-Wave Analysis (RCWA) is mandatory to know how the field reacts to small perturbations in the shaper \citep{RCWA1, RCWA2}. Gratings with small periods generally have some diffracted orders cut off for visible and IR light.

\begin{figure}[!ht]
\begin{center}
\includegraphics[width=9.cm]{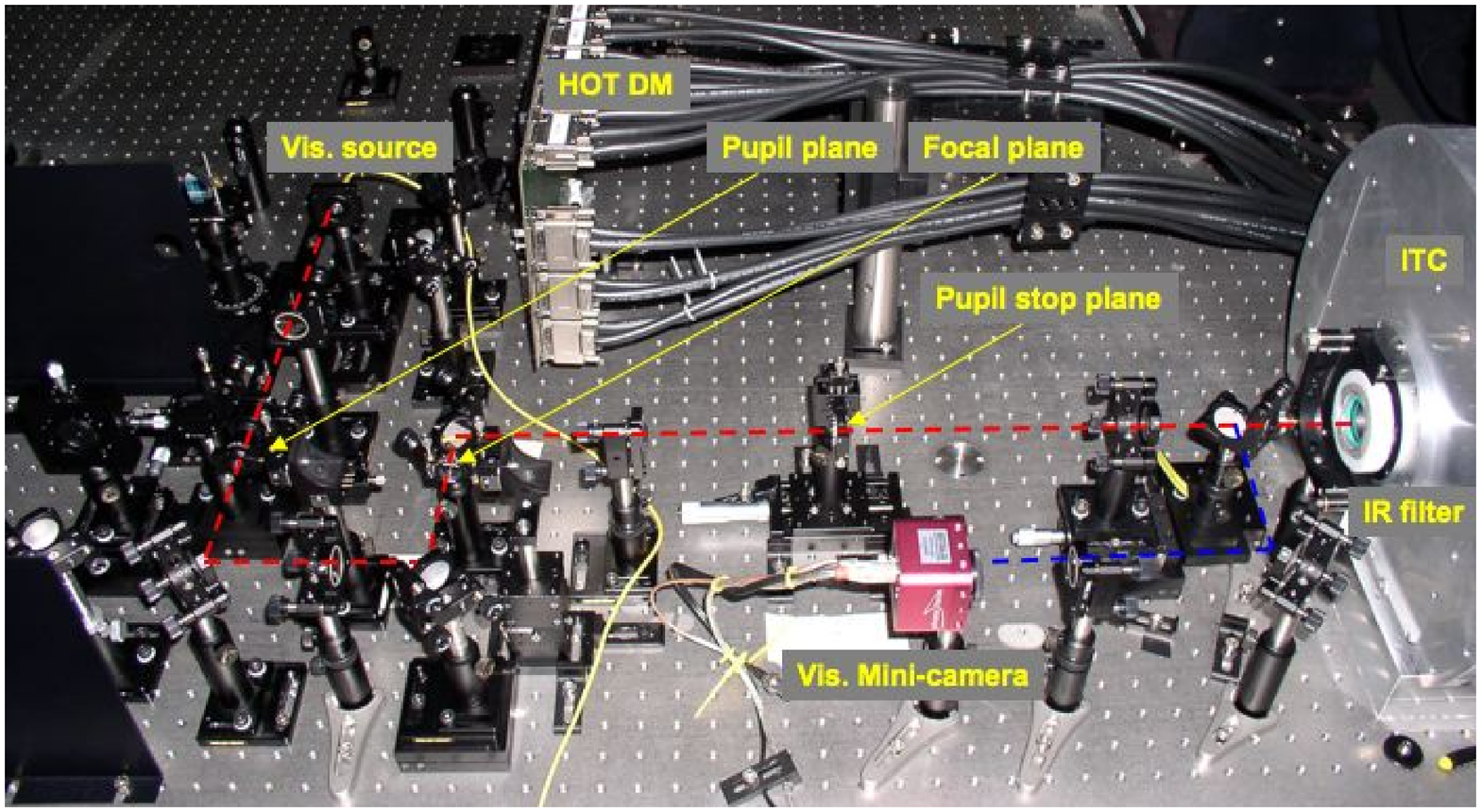}
\includegraphics[width=9.cm]{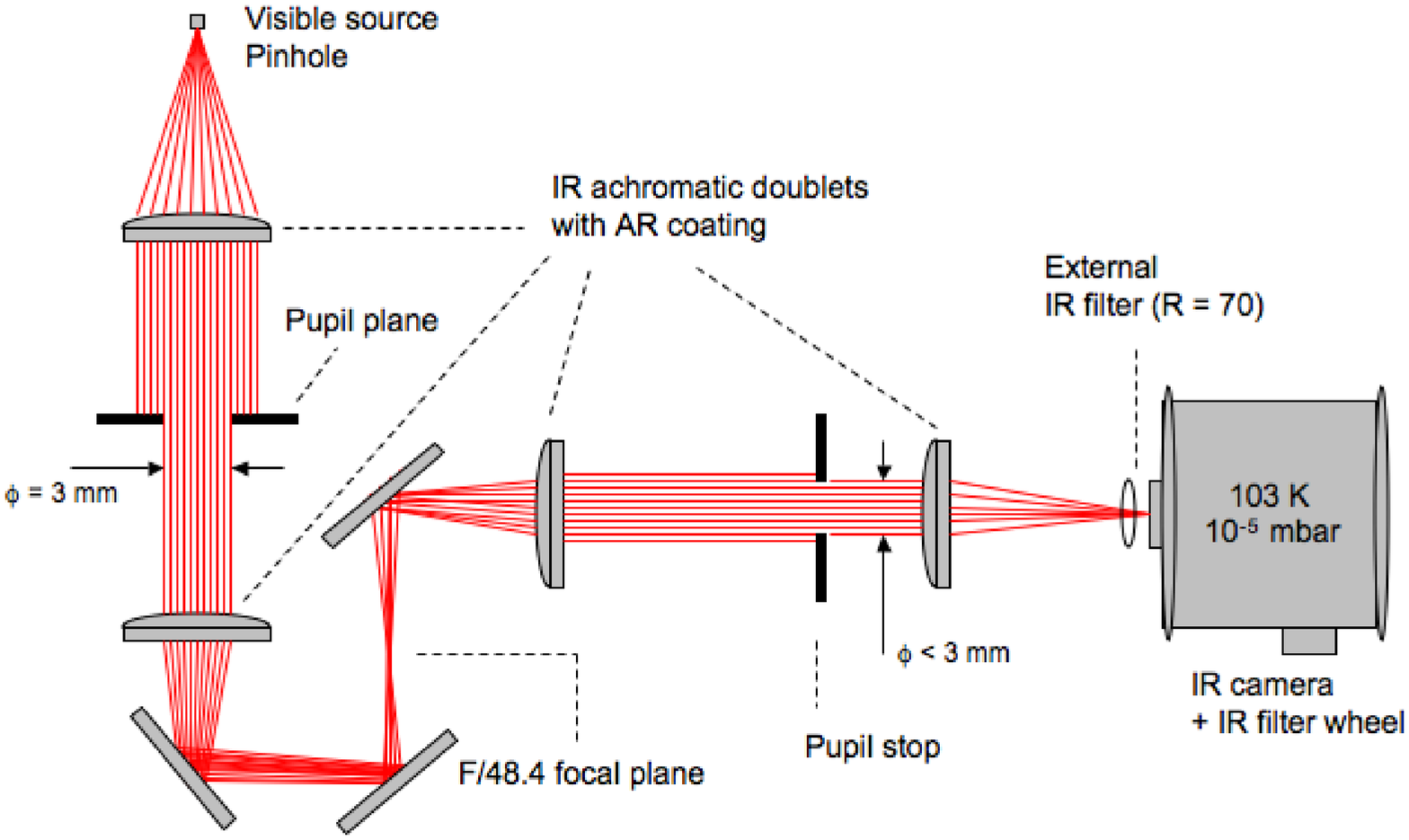}
\end{center}
\caption{$Top$: Picture of the IR coronagraphic test-bench on HOT. The red dot line shows the IR coronagraphic path while the blue dot line shows the pupil imager system path when placing a mirror on a magnetic mount before the external IR filter. 
$Bottom$: shematic setup of the coronagraphic testbench.} 
\label{HOTbench}
\end{figure}

\subsection{Other specifications}
\label{manufacturing}
The microdots apodizer was fabricated by Precision Optical Imaging in Rochester, New York.  
To reduce the effect of misalignment of the apodizer with the  telescope pupil, the designed profile of the apodizer ($\Phi=3mm$) was not obscured at the center by the central obscuration (no 0$\%$ transmission values) and  was extrapolated by a Gaussian function on the outer part (from 1.5 mm to 3 mm in radius, i.e above the apodizer function radius) to slowly decrease the transmission to zero. 
Moreover, having a sharp edge on the apodizer might be detrimental to the characterization process (inspection of the profile), because of strong diffraction effects. 
The shaper was fabricated using wet-etch contact lithography of a Chrome layer (Optical Density of 4.0) deposited on a BK7 glass substrate ($\lambda$/20 peak-to-valley). The back face of the apodizer has an antireflection coating for the H band (1.2 to 1.8 $\mu$m, $R<1\%$).

In the case of wet-ech lithography, etching can lead to a reduction in the light-blocking metal dot sizes (smaller than specified in the digital design), which potentially leads to an increased transmission. 
Dot spacing remains the same, while opaque dot size are reduced in size due to an undercut of the masking layer which form cavities with sloping sidewalls.  
To minimize the impact of this effect on the obtained transmission, the mask design was numerically precompensated by estimating the feature size which would be obtained after fabrication \citep{2007JOSAB..24.1268D}.
In practice, we adopted a pixel grid of 6$\mu m$ (i.e dot size, $S = 500$), and several 
runs were necessary to finely calibrate the process and reach the specification. Reproducibility was confirmed with a last run after optimal conditions were set. 
\begin{figure*}[!ht]
\begin{center}
\includegraphics[width=8.5cm]{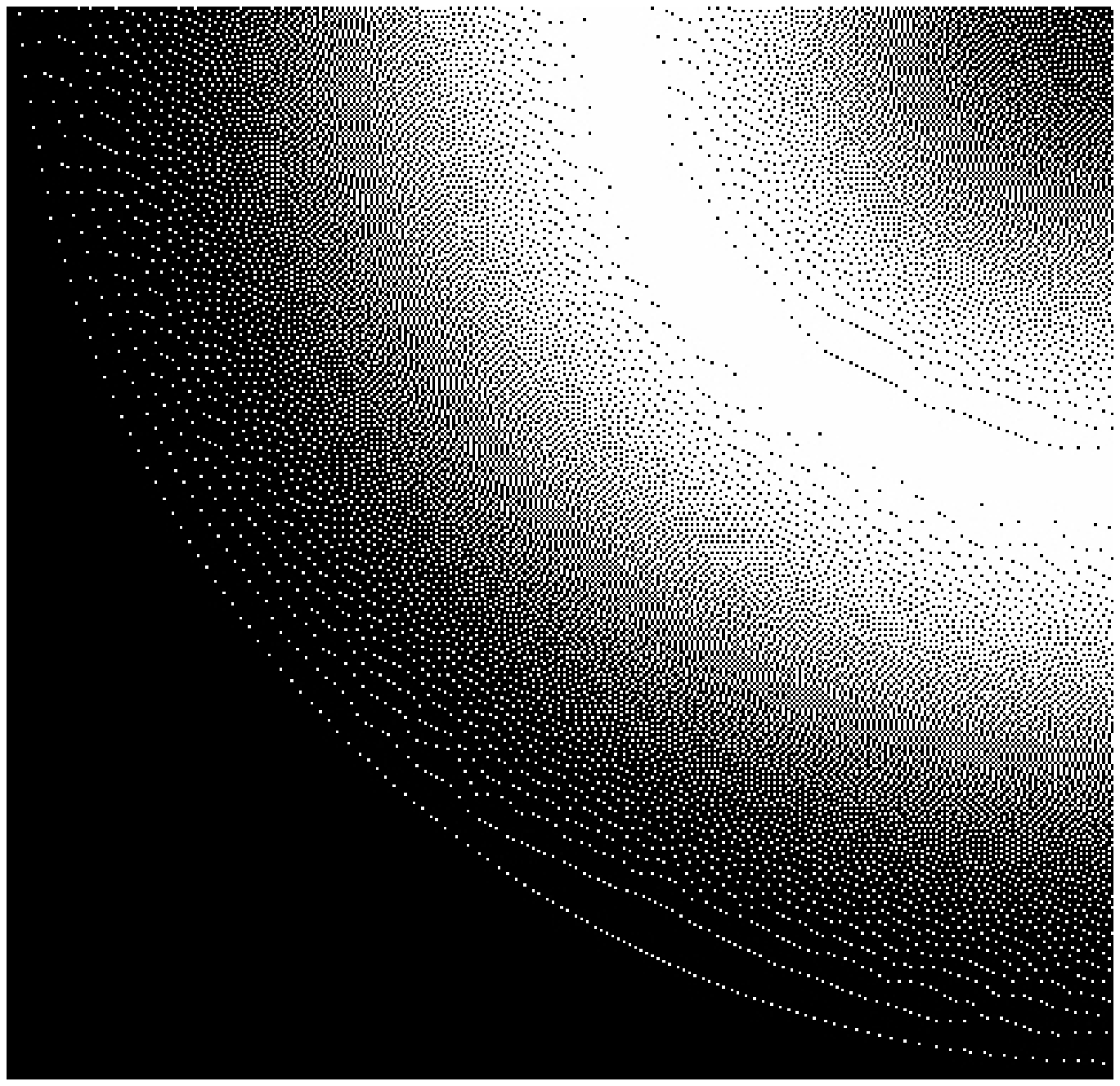}
\includegraphics[width=8.56cm]{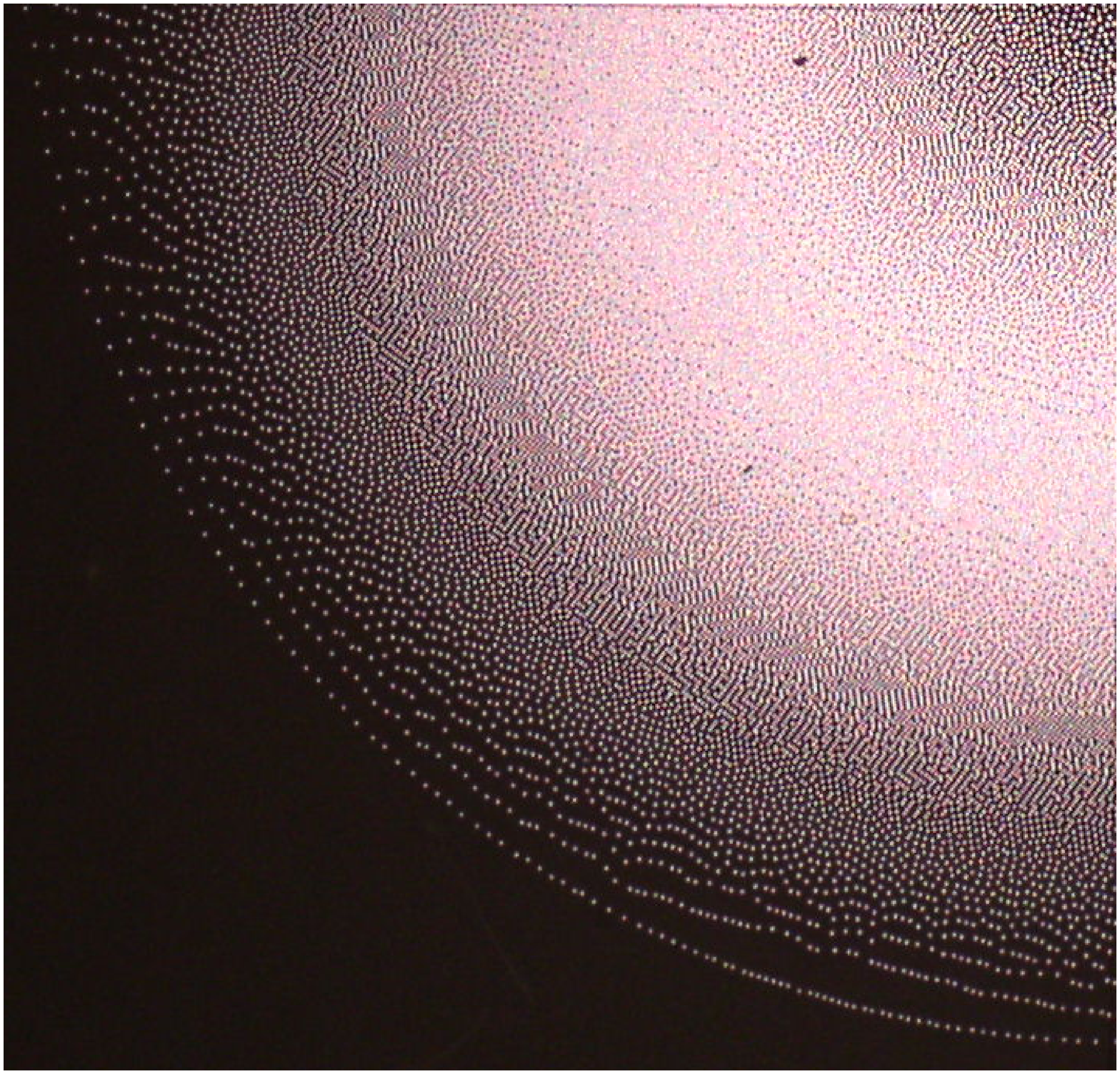}
\end{center}
\caption{Left: simulation map of the binary apodizer with 5$\times$5$\mu$m dots. Right: Shadowgraph inspection of the manufactured microdots apodizer ($\times$50). For the sake of clarity, only a quarter of the apodizer is shown.} 
\label{apod2}
\end{figure*}
\begin{figure*}[!ht]
\centering
\begin{center}
\includegraphics[width=9cm]{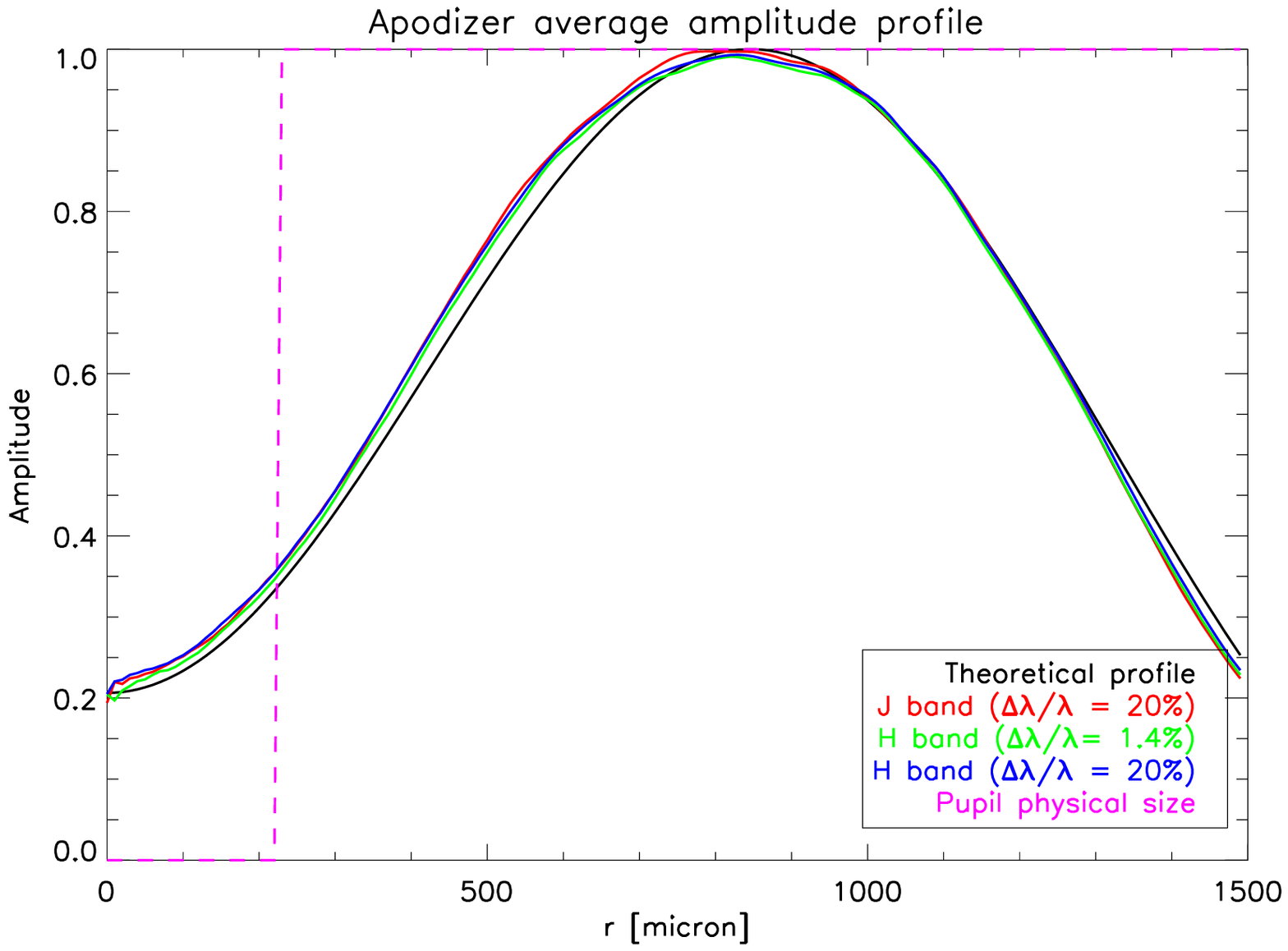}
\includegraphics[width=9cm]{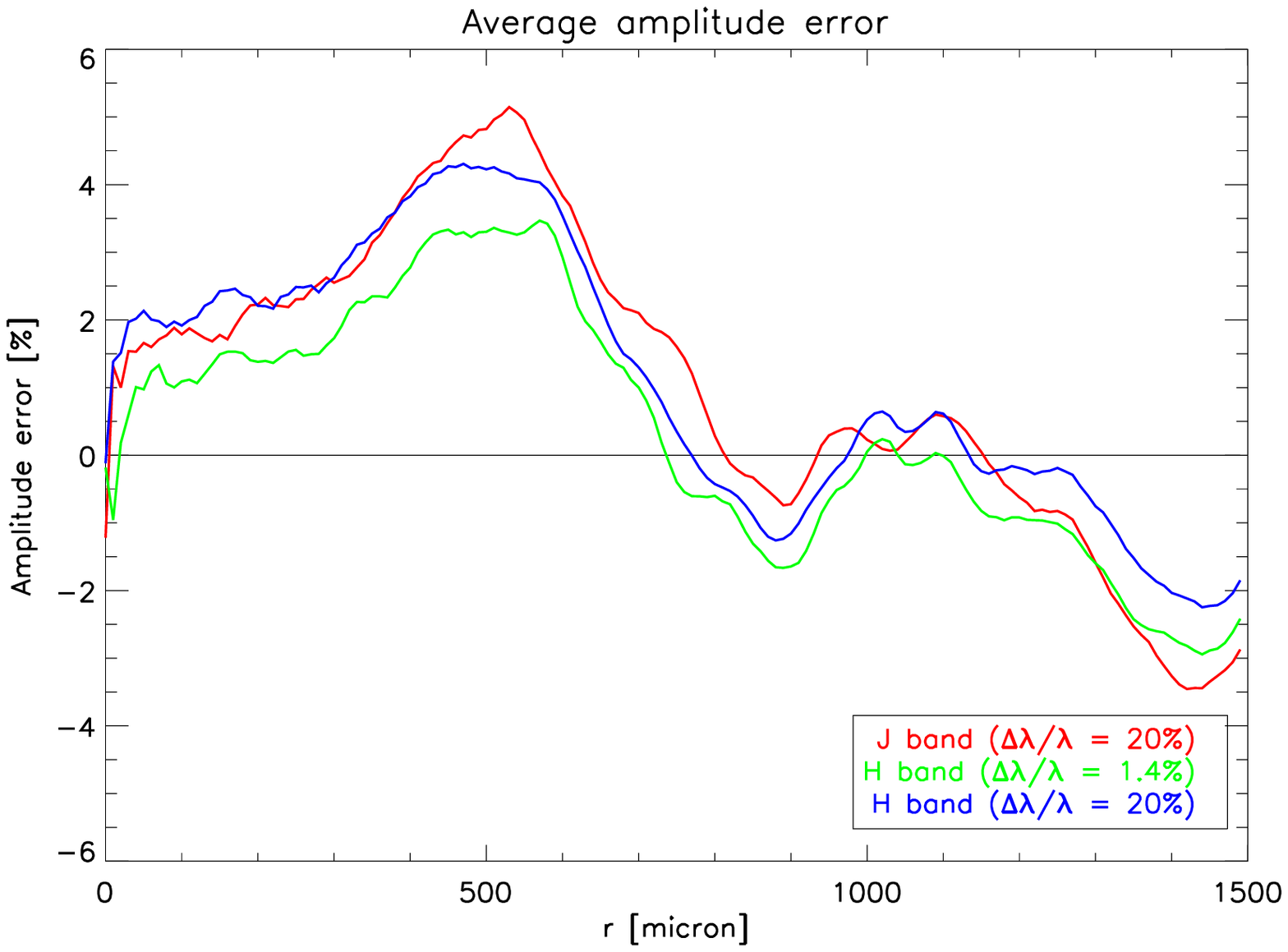}
\end{center}
\caption{Left: Apodizer azimuthally average profile (from center to the edges) using different filters (J, H and narrow H band) compared to specification (black curve). Right: corresponding average amplitude error as function of the position using the same filters.} 
\label{profile}
\end{figure*}  

The 4.5 $\lambda/D$ hard-edge opaque Lyot mask has been fabricated by GEPI, Paris Observatory (360$\mu$m $\pm$ 1$\mu$m in diameter, OD = 6.0 at 1.65 $\mu$m using two metallic layers of Chrome (20 nm) and Gold (200 nm)).


\section{Experiment}
\label{labo}

\subsection{Optical setup}
\label{optical}
The experiment configuration is shown is Fig. \ref{HOTbench}. 
where the optical IR coronagraphic path is described (top) using dot red line on the picture.
The optical setup is designed to simulate the 8 m VLT pupil. The 3mm entrance aperture diameter is made in a laser-cut stainless steel sheet with an accuracy of 0.002 mm.
The central obscuration is scaled to 0.47 mm $\pm$ 0.002 mm and the spider vanes thickness is 15$\mu$m $\pm$ 4$\mu$m. 
\begin{figure}[!ht]
\begin{center}
\includegraphics[width=5.cm]{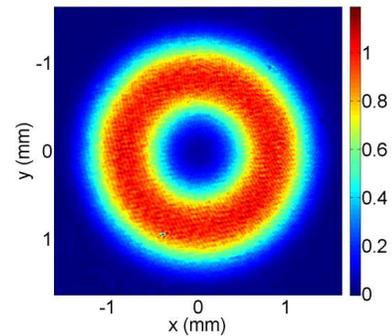}
\end{center}
\caption{Top: Infrared recorded image of the apodizer.} 
\label{apod1}
\end{figure}  
\begin{figure*}[!ht]
\begin{center}
\includegraphics[width=9cm]{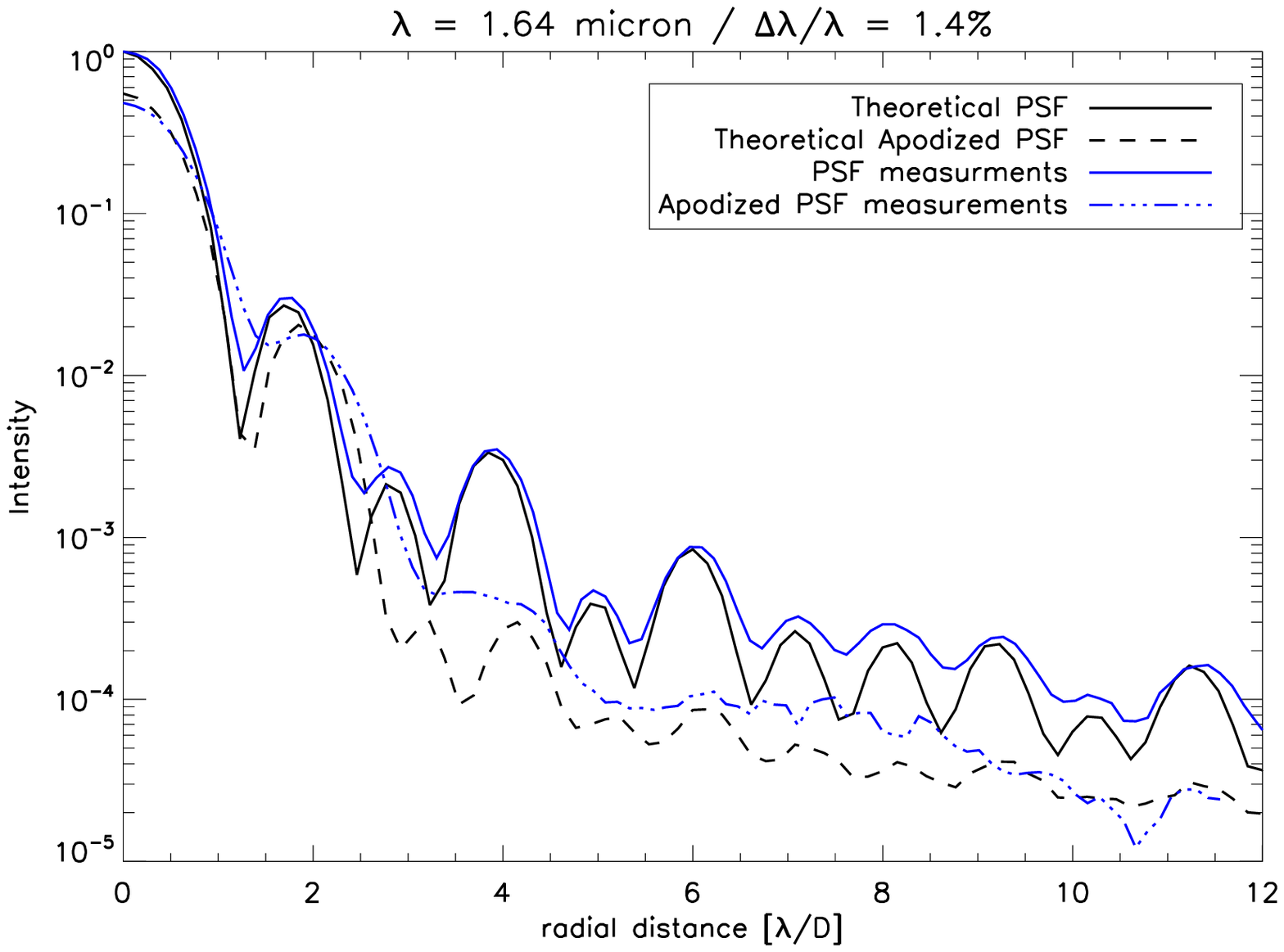}
\includegraphics[width=9cm]{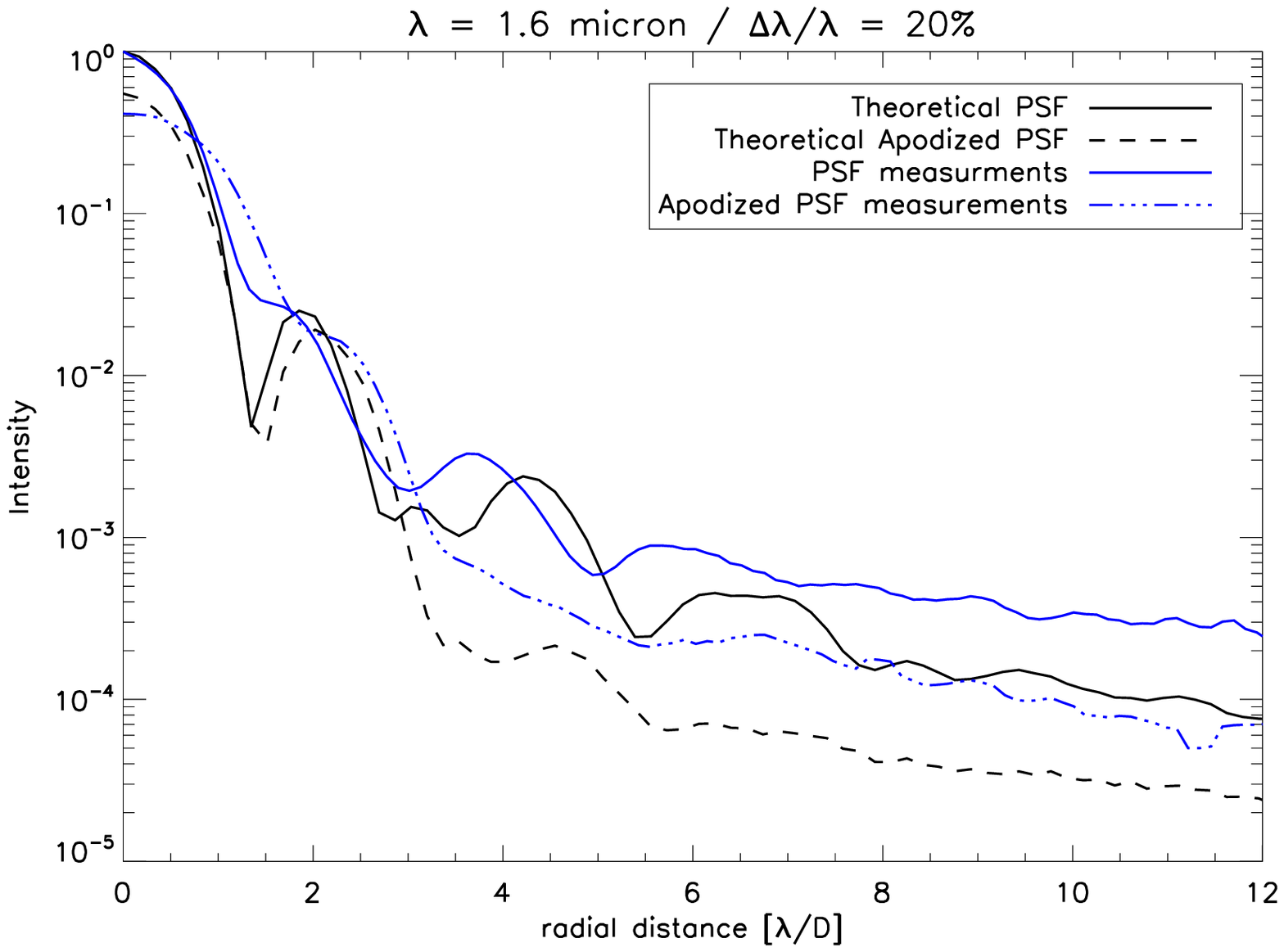}
\end{center}
\caption{Left : PSF and apodized PSF recorded on the bench (blue lines) compared to theoretical ones (black lines) with narrow H filter ($\lambda$= 1.64 $\mu$m, $\Delta \lambda/\lambda$ = 1.4$\%$). Right : Same measurements as previous ones but with broadband H filter ($\Delta \lambda/\lambda$ = 20$\%$).} 
\label{PSF}
\end{figure*} 
The coronagraphic mask is installed at an F/48.4 beam.
Re-imaging optics are made with $\lambda$/10 achromatic IR doublets. 
The quality of the collimation in the pupil plane and re-imaged pupil plane (where the pupil stop is placed) was checked and adjusted using an HASO 64 Shack-Hartmann sensor.

A pupil imager system (see Fig. \ref{HOTbench}, a dot blue line describe its optical path) has been implemented for the alignment of the pupil stop mask with respect to the entrance pupil mask (alignment in x and y direction, orientation of the spider vanes and focalisation as well). 

We installed the entrance pupil mask and the apodizer in the same collimated beam. Hence, the apodizer is not perfectly in the pupil plane.
The apodizer was placed inside a rotating adjustable-length lens tube that allows a translation of $\sim$3.5mm from the pupil mask.

\noindent We used a white-light source combined either with an IR narrow band filter ($\Delta \lambda/\lambda$ = 1.4$\%$, central wavelength of 1.64$\mu$m, with a peak transmission of 64.4$\%$ or IR filters (J, H, K), installed inside the filter wheel of the IR camera where the H filter is centered at 1.6$\mu$m, $\Delta \lambda/\lambda$ = 20$\%$.
The camera used is the ESO Infrared Test Camera (ITC), cooled at 103 K degree with a vacuum of $10^{-5}$ mbar. Internal optics are designed to reach a pixel scale of 5.3 mas. 
The Strehl ratio was evaluated at 94$\%$. 

The APLC pupil stop mimics the VLT pupil mask with spider vanes thickness increased by a factor 4 (60$\mu$m $\pm$ 4$\mu$m), and outer diameter reduced by 0.96$\times \Phi$ (2.88 mm $\pm$ 0.002 mm) and the central obscuration is equal to 0.16$\times \Phi$ (0.49 mm $\pm$ 0.002 mm). The pupil stop throughput is about 90$\%$.

\subsection{Quality of the binary apodizer}
The size of the square chrome dots has been determined to 4.5 $\times$ 4.5 $\mu$m using a microscopic inspection.
The global shape of the binary apodizer is presented in Fig. \ref{apod1}.  The dots spatial distribution across the pupil diameter has been also analyzed using a shadowgraph ($\times$50, Fig. \ref{apod2}) and compared to simulation map (5 $\times$ 5 $\mu$m dots).
Figure \ref{profile} shows that the accuracy on the profile well matches the expected profile, and the transmission error is about 3$\%$. Achromaticity of the profile is also demonstrated : the profile error only increases by 
about 2$\%$ from the narrow H filter to the broadband J filter. 
\begin{figure}[!ht]
\begin{center}
\includegraphics[width=4.3cm]{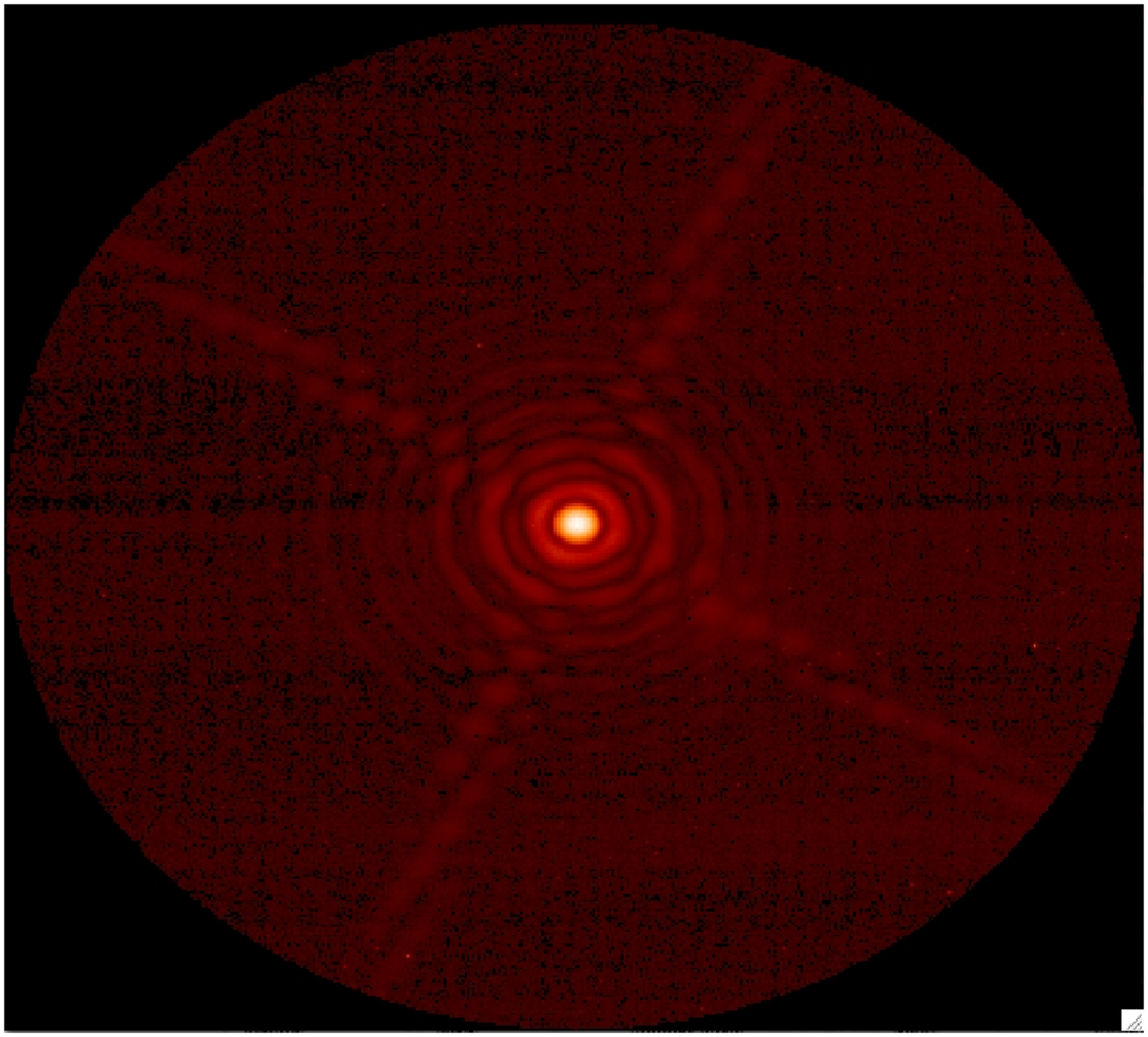}
\includegraphics[width=4.31cm]{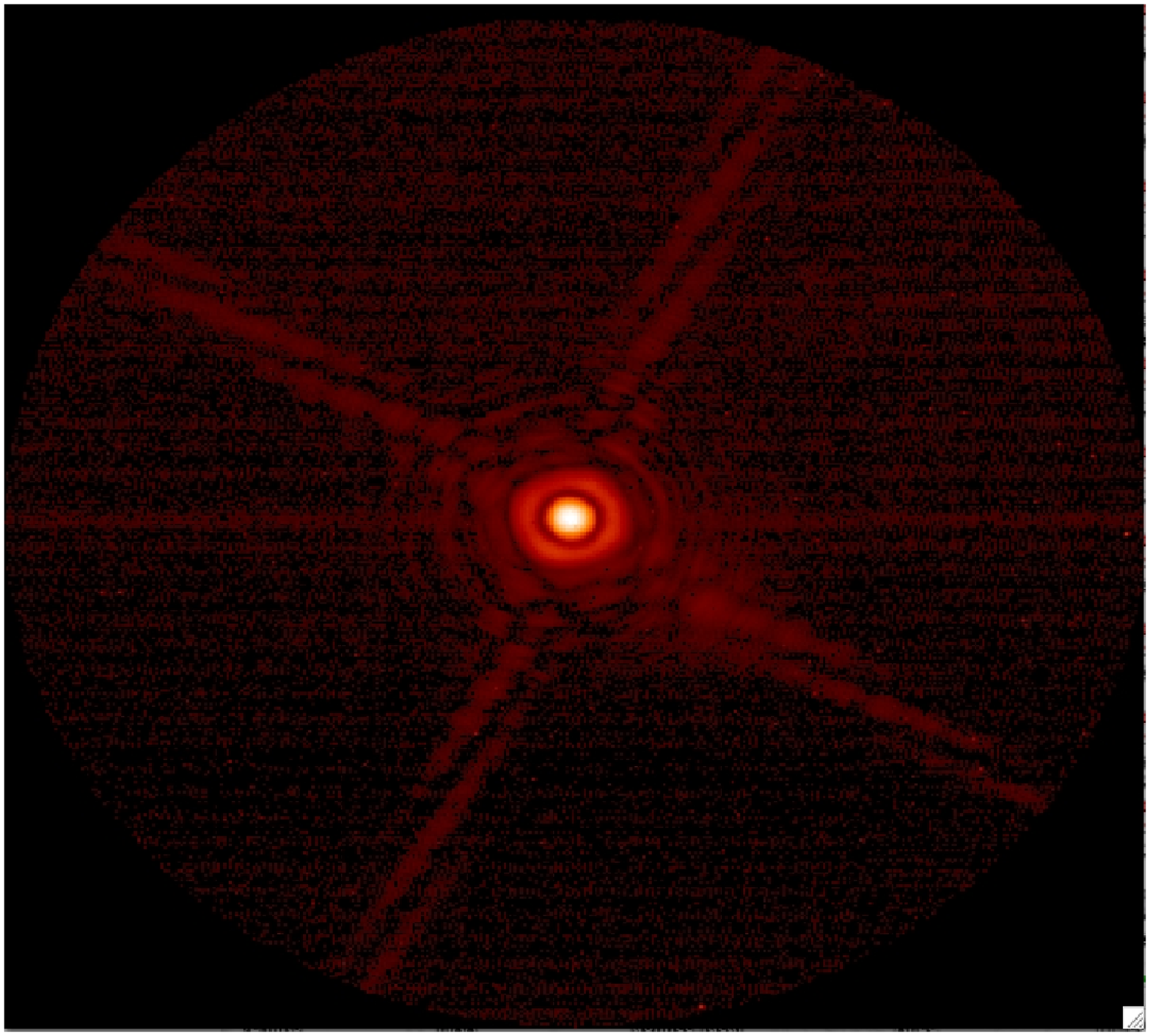}
\end{center}
\caption{Images recorded on the bench ($\lambda$= 1.64 $\mu$m, $\Delta \lambda/\lambda$ = 1.4$\%$), left: VLT-like pupil PSF, right: VLT-like pupil apodized PSF.} 
\label{imPSF}
\end{figure} 
\noindent Having smaller pixel size than the digital design (6$\times$6 $\mu$m) was expected (Sect. \ref{manufacturing}) and demonstrates that precompensation of the transmission error due to the feature size was necessary and works well.

\subsection{Coronagraphic results and discussion}

\subsubsection{Effect on the PSF}
This first series of tests intend to demonstrate the correct behavior of the binary apodizer on the PSF. We only compare the PSF without apodizer to that with the apodizer. Qualitatively (Fig. \ref{imPSF}) it is demonstrated that the apodizer works well : the PSF's wings of the apodized PSF are reduced in intensity while the core of the apodized PSF gets larger (exposure time are here identical, no neutral density is applied). This behavior agrees well with the theoretical predictions.
Although there are some discrepancies between theory and measurements (Fig. \ref{PSF}, bottom, for $\Delta \lambda/\lambda$ = 20$\%$ in H band), the gain between PSF and apodized PSF is consistent with theory. 
This results has been demonstrated in the H-band with a narrow band filter ($\Delta \lambda/\lambda$ = 1.4$\%$) and with a broadband filter ($\Delta \lambda/\lambda$ = 20$\%$). Its achromaticity in H band is therefore confirmed.
The fact that we are using a real optical system and the 3.5mm defocus between the apodizer and the en-
\begin{figure}[!ht]
\begin{center}
\includegraphics[width=5.05cm]{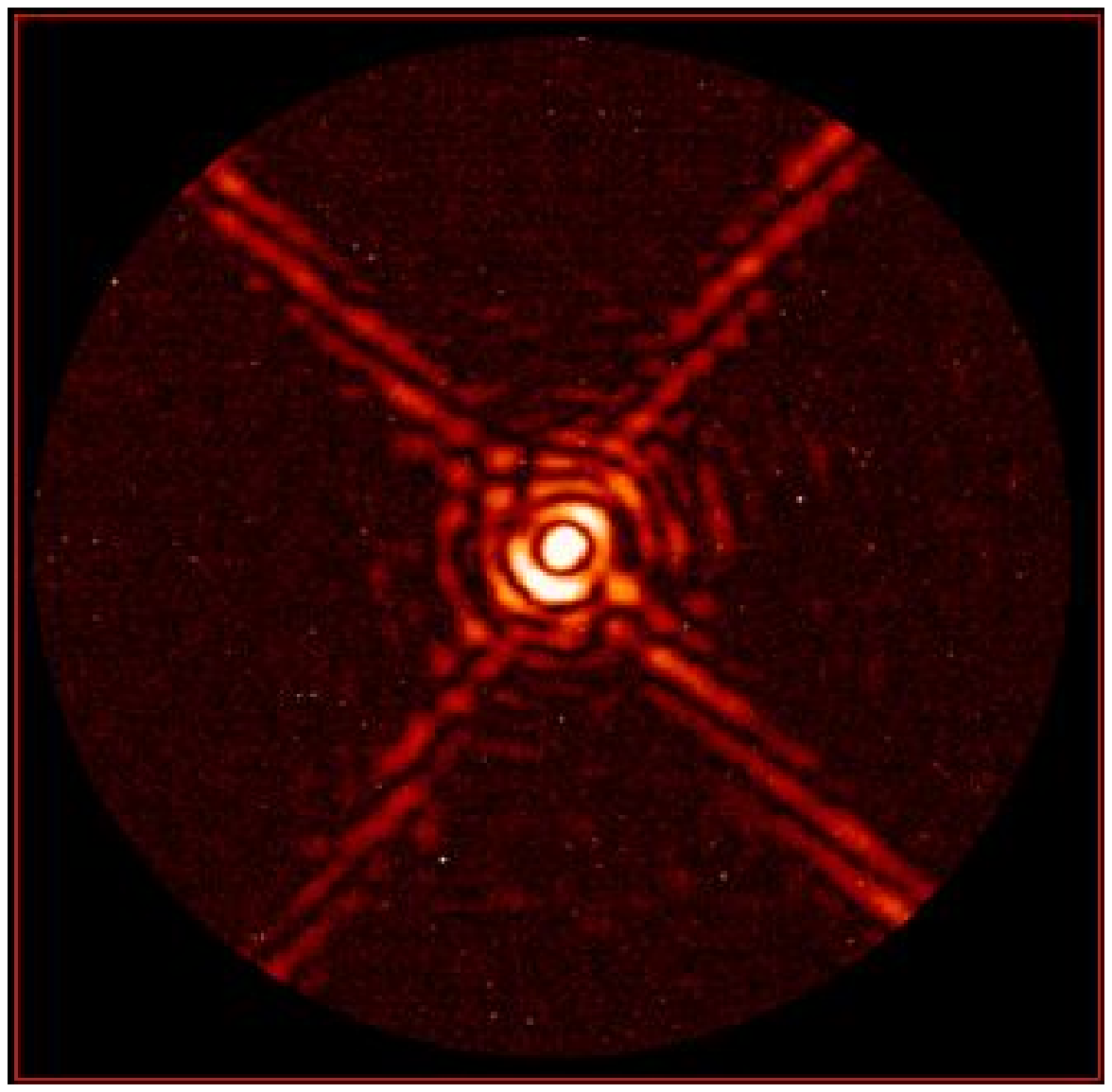}  
\includegraphics[width=1.39cm]{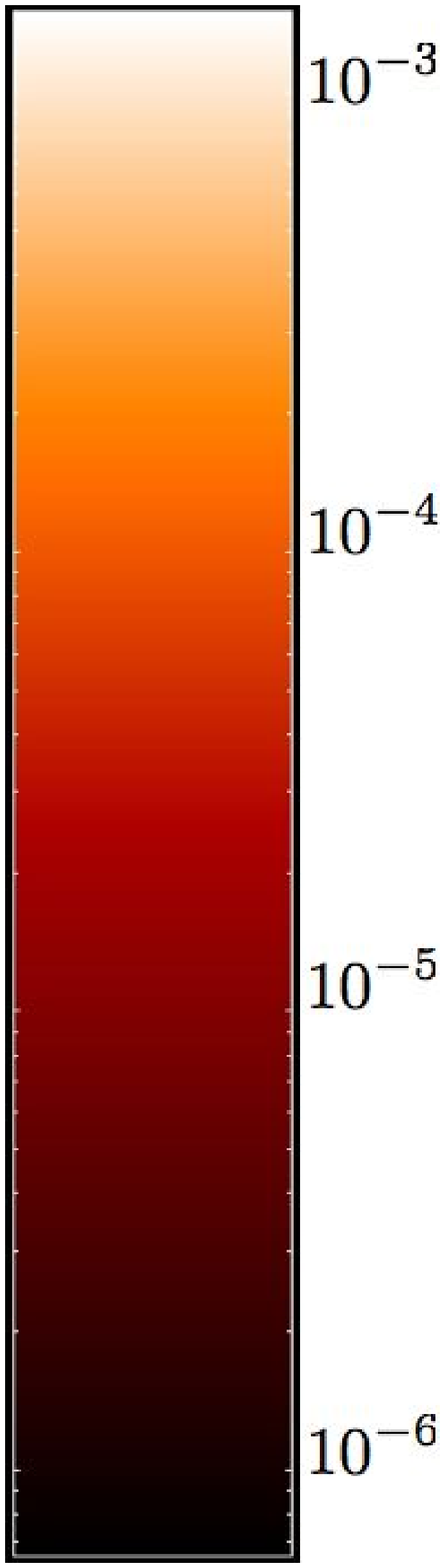}
\end{center}
\caption{Observed raw coronagraphic image (log scale) with its scale of contrast ($\lambda$= 1.64 $\mu$m, $\Delta \lambda/\lambda$ = 1.4$\%$).}
\label{imCORO}
\end{figure} 
\noindent trance pupil may explain the discrepancies. 
\subsubsection{Effect on the coronagraphic PSF}
This second series of tests intend to demonstrate the coronagraphic behavior of the APLC using the microdots apodizer. 

\begin{figure}[!ht]
\begin{center}
\includegraphics[width=9cm]{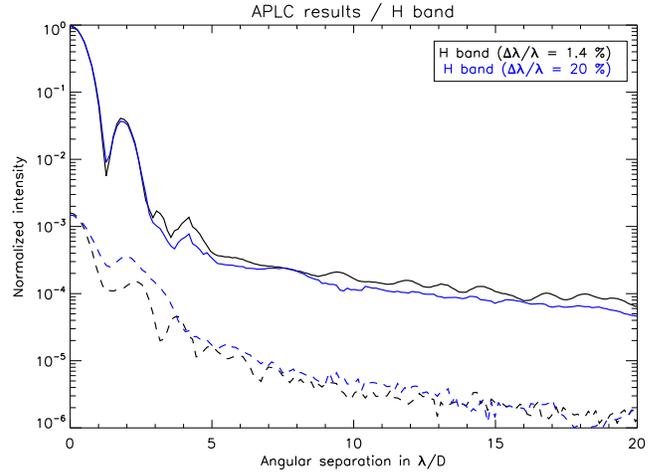}
\end{center}
\caption{Azimuthally averaged coronagraphic profiles at $\lambda$= 1.64 $\mu$m, $\Delta \lambda / \lambda$ = 1.4\% (black lines) and $\Delta \lambda / \lambda$ = 20\% (blue lines).} 
\label{PSF2}
\end{figure}

Qualitatively, the profile of the coronagraphic image (Fig. \ref{imCORO}, H band with $\Delta \lambda/\lambda$ = 1.4$\%$) agrees well with theory: a PSF-like pattern homogeneously reduced in intensity with most of the energy inside the first rings.
In this observed raw image, a local contrast as large as 6.5$\times10^{-7}$ has been reached between the diffraction spikes.
In Fig. \ref{PSF2} we present apodized PSFs and coronagraphic images recorded on the bench using a narrow ($\Delta \lambda/ \lambda = 1.4\%$) and broadband filter ($\Delta \lambda/ \lambda = 20\%$) in the H band.
Most of the time, an order of magnitude discrepancy (mostly in the halo) is found between theory and measured data (Table. \ref{resum}) where we have compared contrast at 3, 12 and 20$\lambda/D$.
The contrast is defined as the ratio of the local intensity (i.e at a given angular separation) on the coronagraphic image to the maximum intensity of the apodized PSF image.
The total rejection rate (ratio between the total intensity of the PSF image and the total intensity of the coronagraphic image, in practice limited to 20$\lambda/D$) is only at a factor of 2 and 1.8 from theory for the narrow and broad band filters respectively.
This discrepancy is reduced when considering the peak rejection (ratio between the maximum intensity of the PSF to the maximum intensity of the coronagraphic image) to a factor of 1.7 and 1.2, respectively.
The impact of chromatism is only slightly revealed at small angular separation (before 4$\lambda/D$), otherwise achromaticity is demonstrated in the halo in H band. 

The discrepancy may find its origin in different error sources (we only discuss here the main ones): 1/ apodizer profile error ($\sim3\%$), 2/ quality of the bench ($Strehl = 94\%$) and 3/ defocus between the apodizer and the pupil plane ($\sim3.5mm$).
Simulations were carried out to analyze independently the impact of the two first ones. 
For the impact of the defocus we refer to a sensitivity analysis performed for SPHERE \citep{Reynard} where the apodizer mask positioning requirement in defocus is set to $\pm0.1mm$. Such error in the positioning impacts mainly the halo.
Including in simulation the measured profile of the apodizer (Fig. \ref{profile}) reduces the discrepancy from 1.7 to 1.2 on the peak rejection and from $\sim$10 to $\sim$3 in the halo.
When the Strehl ratio is set to $94\%$ ($\lambda/25$ nm rms) while the apodizer is perfect, the discrepancy is reduced to 1.08 on the peak rejection and to $\sim$4 in the halo.
It is therefore difficult to conclude on the dominant source of error. The discrepancy with theory is certainly a result of a combination of all theses error sources. 

\begin{center}
\begin{table}[!ht]
\begin{tabular}{l|l|l|l|l}
\hline \hline 
Metrics & \multicolumn{2}{c|}{Measured on bench} & \multicolumn{2}{c}{Theory}  \\
\cline{2-5}
&  \multicolumn{4}{c}{$\Delta \lambda / \lambda [\%] $} \\
\cline{2-5}
&  $1.4$& $20$& $1.4$ & $20$ \\
\hline
\hline
Contrast at 3$\lambda/D$ & 5.0 $10^{-5}$ &  1.5 $10^{-4}$ & 1.4 $10^{-6}$ & 1.2 $10^{-5}$ \\
Contrast at 12$\lambda/D$ &2.3 $10^{-6}$ &  3.5 $10^{-6}$ & 2.1 $10^{-7}$ & 2.8 $10^{-7}$ \\
Contrast at 20$\lambda/D$ &1.2 $10^{-6}$ &  1.8 $10^{-6}$ & 1.0 $10^{-7}$ & 1.3 $10^{-7}$ \\
\hline
Total rejection & 489 & 355 & 1000 & 641 \\
Peak rejection & 627 & 674 & 1058 & 788 \\
\hline
\end{tabular}
\caption{Summary of coronagraphic results and comparison with theory}
\label{resum}
\end{table}
\end{center}
During our laboratory tests, no high frequencies noise due to the apodizer pixellation was revealed. However, simulation analysis presented in Sect. \ref{design} predicts pixellation noise at about 20$\lambda/D$ on the coronagraphic image
at a contrast between $10^{-7}$ and $10^{-8}$ (S = 600). In our case, the contrast level is not deep enough even between the diffraction spikes to reveal the predicted noise.
Therefore, we can only conclude on the performance and suitability of our configuration for HOT (the High Order Testbench developed at ESO, and even for SPHERE) but not on the pixellation noise simultaneously predicted by analytical development (Eq. \ref{principalF} and \ref{I3}) and by simulation. 
We note that smaller pixels size ($<5\mu$m) would certainly help at reducing the 3$\%$ error on the profile which could potentially improve performance.
Despite the discrepancy discussed above, these first results of APLC using microdots apodizer are already beyond the SPHERE requirements \citep{Reynard}.

\section{Conclusion}
\label{conclu}
We report on the development and laboratory experiments of an Apodized Pupil Lyot Coronagraph using microdots apodizer in the near-IR.
Halftone dot process is a promising alternative solution to continuous metal layer deposition. 
Using a diffusion error algorithm, and optimized pixel size and fabrication techniques, we demonstrate impressive agreement between the specified and 
measured transmission profiles, as well as the achromatic behavior of this apodizer.
Coronagraphic properties are consistent with the expected properties, and have already reached the SPHERE requirements. Achromaticity in H band is also demonstrated.

Additionally, pixellated apodizers do not produce a spatially-varying phase aberration which might compromise the coronagraphic effect at all radial distances.

We therefore conclude that microdots apodizers represent a very attractive solution for the APLC. 

Although this study was carried out in a context of Research \& Development for future near IR instruments on E-ELT it is already applicable to other instruments like SPHERE and other coronagraphs like the Dual Zone.   
Finally, we note that a RCWA analysis would be mandatory for a finer analysis of the pixel size with respect to the wavelength.  
\begin{acknowledgements}
P.M would like to thank Sebastien Tordo and Christophe Dupuy from ESO for their helpful support with the ITC and metrology inspection.
This activity is supported by the European Community under its Framework Programme 6, ELT Design Study, Contract No. 011863.
\end{acknowledgements}

\nocite{*}
\bibliography{MyBiblio}
\end{document}